\documentclass[amsmath,amssymb,amsfonts,aps,pre,preprint,superscriptaddress,bibnotes,showpacs,showkeys,longbibliography]{revtex4-1}
\usepackage{mathtools}
\usepackage{epsfig}
\usepackage[english]{babel}
\usepackage{physics}
\usepackage{color}
\usepackage{subcaption}
\usepackage{hyperref}
\usepackage{lipsum}
\usepackage{cases}
\usepackage{bbold}

\newcommand*{\lop}{\mathcal{L}_t}
\newcommand*{\nlop}{\mathcal{N}_t}
\newcommand*{\rop}{\mathcal{R}}
\newcommand{\vect}[1]{\boldsymbol{#1}}
\newcommand{\me}{\mathrm{e}}

\AtBeginDocument{%
    \newwrite\bibnotes
    \def\bibnotesext{Notes.bib}
    \immediate\openout\bibnotes=\jobname\bibnotesext
    \immediate\write\bibnotes{@CONTROL{REVTEX41Control}}
    \immediate\write\bibnotes{@CONTROL{%
    apsrev41Control,author="08",editor="1",pages="1",title="0",year="1"}}
     \if@filesw
     \immediate\write\@auxout{\string\citation{apsrev41Control}}%
    \fi
}%

\begin{document}
\title{Epidemic processes with vaccination and immunity loss studied with the BLUES function method.}

\author{Jonas Berx}
\affiliation{Institute for  Theoretical Physics, KU Leuven, B-3001 Leuven, Belgium}

\author{Joseph O. Indekeu\footnote{Given his role as Editor of this journal, Joseph O. Indekeu had no involvement in the peer-review of articles for which he was an author and had no access to information regarding their peer-review. Full responsibility for the peer-review process for this article was delegated to another Editor.}}
\affiliation{Institute for  Theoretical Physics, KU Leuven, B-3001 Leuven, Belgium}

\date{\today}

\begin{abstract}
The Beyond-Linear-Use-of-Equation-Superposition (BLUES) function method is extended to coupled nonlinear ordinary differential equations and applied to the epidemiological SIRS model with vaccination. Accurate analytic approximations are obtained for the time evolution of the susceptible and infected population fractions. The results are compared with those obtained with alternative methods, notably Adomian decomposition, variational iteration and homotopy perturbation. In contrast with these methods, the BLUES iteration converges rapidly, globally, and captures the exact asymptotic behavior for long times. The time of the infection peak is calculated using the BLUES approximants and the results are compared with numerical solutions, which indicate that the method is able to generate useful analytic expressions that coincide with the (numerically) exact ones already for a small number of iterations.

\end{abstract}

\maketitle

\section{Introduction}
\label{sec:introduction}

Systems of coupled differential equations (DEs) are omnipresent in the sciences, medicine and even in the humanities. While some of these systems have exact closed-form solutions, the majority do not and hence need to be solved by numerical means. One way to supplement direct numerical simulation with useful information about the (physical) structure behind the coupled DEs is to use approximate (semi-)analytical solutions, in which the ``physical'' role played by the parameters of the DE is conspicuous.

A stronger need for analytical techniques arises when the parameter space to be scanned is large. For each set of parameters the numerical computation must be repeated, which, for large phase spaces, is computationally expensive. Hence, analytical solutions or approximations are of significant value in this context. Furthermore, numerical techniques can often be unreliable as a result of rounding errors, discretization, etc. For specific problems, one usually requires a specific numerical algorithm that leads to useful results. (Semi-)analytical methods do not rely on the use of discretization, perturbation or linearization and can handle most types of nonlinearities where numerical solvers become very complicated, e.g., integro-differential equations or fractional differential equations.

The best-known examples of methods to calculate such analytical approximants are the Adomian decomposition method (ADM) \cite{adomian,ADOMIAN199017}, the variational iteration method (VIM) \cite{HE20073}  and the homotopy perturbation method (HPM) \cite{HE1999257}. However, a drawback of these methods is the poor convergence for large values of the independent variable (e.g., for long times). 

A well-known example of nonlinear coupled DEs is the susceptible-infected-recovered-susceptible (SIRS) model for studying the time-evolution of infectious diseases \cite{sir}.  There have been attempts to generate approximate solutions for this model using the ADM \cite{MAKINDE2007842}, the VIM and the HPM \cite{ghotbi2011,MUNGKASI20211} but these solutions diverge quickly and necessitate an improvement of the methods in order to extend the region of convergence.

This paper is structured as follows: in Section \ref{sec:BLUES_method_system} we extend the by now established BLUES function method \cite{Indekeu_2018,berx,Berx_2020,Berx2021_PDE} to a system of nonlinear coupled ordinary DEs. Next, in Section \ref{sec:SIRS}, we describe the SIRS model with a constant vaccination strategy and reduce the system to the two-dimensional subsystem we subsequently solve. In Section \ref{sec:BLUES} we set up the BLUES function method for coupled DEs and apply it, generating approximate solutions to the SIRS model. We stress the importance of the identification of the steady states to construct the associated linear operator. The results are compared with the numerical solution and with results from three other methods, being the ADM, VIM and HPM. We use the analytical approximants to calculate the time at which the infection peak occurs and compare with numerical results. Finally, in Section \ref{sec:conclusions} we present the conclusions and an outlook.

\section{BLUES function method for coupled differential equations}\label{sec:BLUES_method_system}

Here we extend the BLUES iteration originally developed for ordinary DEs \cite{Indekeu_2018, berx,Berx_2020} and partial DEs \cite{Berx2021_PDE} to a system of coupled ordinary DEs. The role of the inhomogeneous {\em source (or sink) term} in the context of the ordinary DE will now be taken over by a vector of sources (or sinks). 

Let us start from an $n$-dimensional system of inhomogeneous nonlinear coupled ordinary DEs that can be written as a nonlinear operator $\nlop$ acting on a vector $\vect{X}(t)$, with source vector $\vect{\chi}(t)$,
\begin{equation}
    \label{eq:nonlinear_operator_chi}
    \nlop \vect{X}(t) = \vect{\chi}(t),\; \forall t > 0
\end{equation}
We now judiciously decompose the nonlinear operator $\nlop$ into a linear operator $\lop$, which contains {\em inter alia} a first derivative in time, and no higher time derivatives, and a residual operator $\rop$, i.e., $\rop \equiv \lop - \nlop$, which contains the nonlinear part of $\nlop$. We add the subscript $t$ to the linear and nonlinear operators to emphasize their dependence on time. Defining suitable initial conditions for $t=0$, i.e.,
\begin{equation}
    \label{eq:boundary_conditions}
   \vect{X}(0) = \vect{C}\,,
\end{equation}
completes the description of the system. In the application we consider in this work, $\rop$ does not depend on $t$. Thus the action of the linear operator on $\vect{X}$ results in the following associated linear coupled system
\begin{equation}
    \label{eq:linear_operator}
    \lop\vect{X} =  \vect{X}_t - A \vect{X} = \vect{\chi},\; \forall t >0
\end{equation}
where the subscript denotes derivative w.r.t. time and the same boundary conditions are imposed as in  \eqref{eq:nonlinear_operator_chi}. The elements of the matrix $A$ are constants in most of the applications we have in mind. We now propose to rewrite the system of DEs in an equivalent form by incorporating the initial condition  through multiplication of $\vect{C}$ with a Dirac delta source $\delta(t)$ located at $t=0$ and by including this term on the right-hand-side of the inhomogeneous system, i.e., 
\begin{equation}
    \label{eq:linear_operator_dirac}
    \lop\vect{X} =  \vect{X}_t - A \vect{X} = \vect{\chi} +\vect{C} \delta \equiv \vect{\psi}, \;\forall t\geq0
\end{equation}
where we have combined the external source $\vect{\chi}$ and the ``initial condition source" $\vect{C}\delta$ into the combined source $\vect{\psi}$. This formulation amounts to resetting the initial condition to zero, so that $ \vect{X}(t) =0$ for $t \leq 0^-$, followed by a jump in $\vect{X} $ implied by integrating the DE over the delta source, so that $ \vect{X}(0^+) = \vect{C}$ and $\vect{X}(t)$ evolves in a continuous manner for $t>0$. The solution of this linear system \eqref{eq:linear_operator_dirac} is the following convolution integral \cite{Zozulya2017,Godunov}
\begin{equation}
    \label{eq:linear_solution}
    \vect{X}(t) = (G\ast\vect{\psi})(t) = G(t)\vect{C} + \int_\mathbb{R} G(t-t')\vect{\chi}(t')\mathrm{d}t',\;\;\forall t > 0
\end{equation}
where $G(t)$ is the Green function matrix for the inhomogeneous linear system. This object can be calculated by finding the matrix exponential $\exp(At) \equiv \mathbb{1} + At + ...$, i.e.,
\begin{equation}
    \label{eq:matrix_exponential}
    G(t) = \me^{At}\Theta(t)\,.
\end{equation}
with $\Theta(t)$ the Heaviside step function (which we take to be unity for positive times {\em including} $t=0$). This Green function matrix solves the linear system with a delta function unit matrix source, i.e., it is a solution of the matrix equation
\begin{equation}
    \label{eq:linear_solve_green}
    G_t - A G = \delta(t) \mathbb{1}, \; \forall t\geq0.
    \end{equation}

Adopting the BLUES function strategy, a solution to the nonlinear system \eqref{eq:nonlinear_operator_chi}, rewritten in the equivalent form 
\begin{equation}
    \label{eq:nonlinear_operator_dirac}
    \nlop\vect{X} =   \vect{\chi}+\vect{C} \delta\equiv \vect{\psi}, \;\forall t\geq0
\end{equation}
is now proposed in the form of a convolution $\vect{X}(t) = (B\ast\vect{\phi})(t)$, in which the function $B(t)$, named BLUES function, is taken to be equal to the Green function of the chosen related linear system, i.e., $B(t) = G(t)$ and the new  source $\vect{\phi}(t)$ is to  be calculated by systematic iteration, using the given (combined) source $\vect{\psi}(t)$. This procedure starts from the following implicit equation, which makes use of the action of the residual operator,
\begin{equation}
    \label{eq:residual_operator_action}
    \begin{split}
     \vect{\phi} &=   \vect{\psi} + (\vect{\phi} -      \vect{\psi}) \\    &= \vect{\psi} +  \rop (B\ast\vect{\phi}).
    \end{split}
\end{equation}

To find the solution to the nonlinear system \eqref{eq:nonlinear_operator_chi}, equation \eqref{eq:residual_operator_action} can be iterated to calculate an approximation for $\vect{\phi}$ in the form of a sequence in powers of the residual $\rop$. This leads to the sequence $\vect{\phi}^{(n)}(t)$, with $\vect{\phi}^{(0)}(t) = \vect{\psi}$. By subsequently taking the convolution product with $B(t)$, approximate solutions $\vect{X}_\psi^{(n)}(t)$ to \eqref{eq:nonlinear_operator_chi} can be obtained. Explicitly, the iteration for these approximants reads 
\begin{equation}
    \label{eq:nth_order}
    \vect{X}^{(n)}_\psi(t) = (B\ast\vect{\phi}^{(n)})(t) = \vect{X}^{(0)}_\psi(t) + \left(B\ast\rop  \vect{X}^{(n-1)}_\psi\right)(t)\, ,
\end{equation}
where
\begin{equation}
    \label{eq:zeroth_order}
    \vect{X}^{(0)}_\psi(t) = (B\ast\vect{\phi}^{(0)})(t)= (B\ast\vect{\psi})(t)
\end{equation}
is the zeroth approximant, which is the convolution product of the linear problem. We now turn to applying the BLUES iteration procedure to the  SIRS epidemiological model for the spreading of infectious diseases. 

\section{The SIRS model with constant vaccination}
\label{sec:SIRS}
\begin{figure}[!htp]
    \centering
    \includegraphics[width=0.85\linewidth]{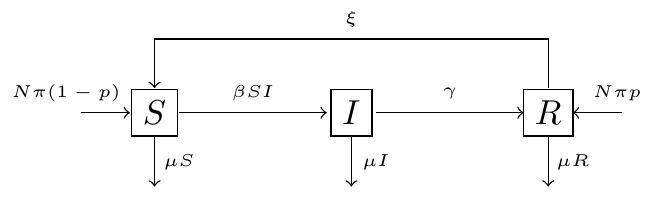}
    \caption{SIRS model with vaccination, immunity loss, births and natural deaths.}
    \label{fig:sirs_chart}
\end{figure}

The SIRS model consists of a group of susceptible $(S)$, infected ($I$) and recovered/immune ($R$) (human) individuals. The total population $N = S+I+R$ can grow by virtue of a (constant) birth rate $\pi$ and decay through natural deaths at a rate $\mu$. At birth, the individuals are vaccinated with probability $p$ and consequently acquire immunity, effectively adding up to the group of immune individuals. The remainder of the new births are added to the susceptible group with probability $1-p$. Through contact with infected individuals, a person can become infected, following a mass-action law $\beta SI$ with force of infection $\beta I$. The infected can recover with rate $\gamma$, acquiring (temporary) immunity and moving to the group of recovered or immune individuals. Finally, immunity can be lost with rate $\xi$, whereby recovered individuals move back to the susceptible population. Fig.\ref{fig:sirs_chart} illustrates these different processes in a flow diagram. We assume that all system parameters and populations are positive, and $0 \leq p \leq 1$.

The interactions between the different populations can be described by the following nonlinear system of DEs, in which the prime denotes derivative w.r.t. time,
\begin{subequations}
\label{eq:SIR}
\begin{align}
S'(t) &=  N(t)\pi(1-p) -\beta \frac{S(t)I(t)}{N(t)} - \mu S(t) +\xi R(t) \label{eq:SIRS_S}\\
I'(t) & = \beta \frac{S(t)I(t)}{N(t)}-(\gamma+\mu)I(t) \label{eq:SIRS_I}\\
R'(t) & =  N(t)\pi p +\gamma I(t)-(\mu+\xi) R(t).\label{eq:SIRS_R}
\end{align}
\end{subequations}
The time evolution of the  total population $N=S+I+R$ can be found by adding \eqref{eq:SIRS_S}-\eqref{eq:SIRS_R}
\begin{equation}
    N'(t) = (\pi-\mu)N(t)\, ,
\end{equation}
which indicates that the population is not constant. $N(t)$ is a nondecreasing function of $t$ when the birth rate is higher than or equal to the death rate, $\pi\geq\mu$. To study the relative importance of the various population fractions, we scale $S,I$ and $R$ by the total population $N$, i.e., $s(t) = S(t)/N(t)$, $i(t) = I(t)/N(t)$ and $r(t)=R(t)/N(t)$. This transforms the system \eqref{eq:SIR} into the following system for the fractions,
\begin{subequations}
\label{eq:SIR_normalized}
\begin{align}
s'(t) &=  \pi(1-p) -\beta s(t)i(t) - \pi s(t) +\xi r(t) \label{eq:SIRS_s}\\
i'(t) & = \beta s(t)i(t)-(\gamma+\pi)i(t) \label{eq:SIRS_i}\\
r'(t) & =  \pi p +\gamma i(t)-(\pi+\xi) r(t)\,,\label{eq:SIRS_r}
\end{align}
\end{subequations}
where $s(t) + i(t) + r(t) = 1$, $\forall t\geq0$. Note that $\mu$ is eliminated by this transformation. By using the constraint on the population fractions, we can eliminate $r(t)$ and study the ``two-dimensional" invariant system
\begin{subequations}
\label{eq:SIR_normalized_invariant}
\begin{align}
s'(t) &=  \pi(1-p) -\beta s(t)i(t) - (\pi+\xi) s(t) -\xi i(t) +\xi\label{eq:SIRS_invariant_s}\\
i'(t) & = \beta s(t)i(t)-(\pi +\gamma)i(t)\, . \label{eq:SIRS_invariant_i}
\end{align}
\end{subequations}
From a stability analysis performed in appendix \ref{sec:stability}, we deduce that the above system has two globally stable fixed points, which are known exactly: a disease-free equilibrium \eqref{eq:fixed_points_0} $\varepsilon_0 \equiv (s^*_0,i^*_0) =\left(1-\frac{\pi p}{\pi +\xi},0\right)$ for which the disease is eradicated and an endemic equilibrium \eqref{eq:fixed_points_e_full} $\varepsilon_e \equiv (s^*_e,i^*_e)= \left(\frac{\pi+\gamma}{\beta},\frac{\beta((1-p)\pi +\xi) -(\gamma+\pi)(\xi+\pi)}{\beta(\gamma+\pi+\xi)}\right)$ for which the disease persists and keeps circulating through the population. The final state of the system is characterized by the \emph{vaccination reproduction number $R_V$}
\begin{equation}
    \label{eq:R0}
    R_V = \frac{\beta\left((1-p)\pi +\xi\right)}{(\pi+\gamma)(\pi+\xi)}\,,
\end{equation}
which represents the average number of susceptible individuals which are infected by one sick individual  during their infectious period, while a vaccination program is in long-time use \cite{VANDENDRIESSCHE2017288}.

Obviously, the endemic equilibrium $\varepsilon_e$ can only exist when $s^*_e<1$ and $i^*_e>0$ which means that the vaccination reproduction number must satisfy $R_V>1$. The disease will be fully eradicated whenever the disease-free equilibrium is the only possible stable fixed point and the endemic equilibrium does not exist. This happens when the vaccination probability $p$ is higher than the critical vaccination threshold $p_c$, which can be inferred from equation \eqref{eq:R0} by setting $R_V=1$,  
\begin{equation}
    \label{eq:sirs_pc}
    p_c = \left(\frac{\xi}{\pi}+1\right)\left(1-\frac{\gamma+\pi}{\beta}\right).
\end{equation}

Note that $p_c$ may exceed unity, whereas $p$ cannot. Note that one could introduce an active constant vaccination strategy which would amount to moving individuals from $S$ to $R$ with a rate $\omega$. This introduces extra terms into the nonlinear system \eqref{eq:SIR} and results in transforming the term $(\pi+\xi) s(t)$ into $(\pi+\xi+\omega) s(t)$ in the $s-$channel of the reduced subsystem \eqref{eq:SIR_normalized_invariant}. The vaccination reproduction number $R_V(\omega)$ in this situation would then be equal to
\begin{equation}
    \label{eq:R0_omega}
    R_V(\omega) = \frac{\beta\left((1-p)\pi +\xi\right)}{(\pi+\gamma)(\pi+\xi+\omega)}\,,
\end{equation}
which amounts to rescaling the $\omega = 0$ case, i.e.,
\begin{equation}
    \label{eq:R0_omega_relation}
    R_V(\omega) = \frac{\pi+\xi}{\pi+\xi+\omega}\, R_V(0)\,.
\end{equation}
It is now clear that $\forall\omega>0$, i.e., active vaccination, $R_V(\omega)<R_V(0)$. Hence, active vaccination acts as an additional mechanism to decrease the reproduction number and possibly to suppress the existence of an endemic equilibrium. For the sake of simplicity, we will choose $\omega = 0$ in this work.

\section{BLUES function method for the SIRS model}\label{sec:BLUES}
To find solutions of the system \eqref{eq:SIR_normalized_invariant}, we can write it as a nonlinear matrix equation, as was demonstrated in section \ref{sec:BLUES_method_system}, i.e., 
\begin{equation}
    \label{eq:matrix_ODE}
    \nlop\vect{X}(t) = \vect{\psi}(t)
\end{equation}
with $\vect{X}(t)$ the vector of solutions,
\begin{equation}
    \label{eq:vector_solutions}
    \vect{X}(t) = \begin{pmatrix}
        s(t) \\
        i(t) \\
    \end{pmatrix}
\end{equation}
and with source vector $\vect{\psi}(t) = \vect{\chi}\ + \vect{C}\delta(t)$, with $\vect{\chi}$  the (generally time-dependent) vector of external sources and $\vect{C}$ the vector of initial conditions, i.e.,
\begin{equation}
    \label{eq:source}
    \vect{\chi} = \begin{pmatrix}
        \chi_s \\
        \chi_i \\
    \end{pmatrix} \qquad \; \mbox{and} \; \qquad \vect{C} = \begin{pmatrix}
        s_0 \\
        i_0 \\
    \end{pmatrix} \equiv \begin{pmatrix}
        s(0) \\
        i(0) \\
    \end{pmatrix}\, .
\end{equation}
In this work we will choose a constant vaccination strategy at birth, i.e., $\chi_s$ and $\chi_i$ are time-independent. Note that we have included the initial conditions in the source $\vect{\psi}$ by multiplication with a Dirac point source located at $t=0$, as was explained in Section \ref{sec:BLUES_method_system}.

Now we judiciously tailor the linear operator that is congruous with the asymptotic equilibrium, by rewriting the nonlinear term in \eqref{eq:SIR_normalized_invariant} so that already the linear system captures the stable fixed point exactly. This is done by including the deviations of the population fractions from their equilibrium values in the (revised) nonlinear term, as follows,
\begin{subequations}
\label{eq:SIR_FPexpanded}
\begin{align}
s'(t) &=  \pi(1-p) -\beta (s(t)-s^*)(i(t)-i^*) - (\pi+\xi-\beta i^*) s(t) -(\xi+\beta s^*) i(t) +\xi + \beta s^*i^* \label{eq:SIRS_FPexpanded_s}\\
i'(t) & = \beta (s(t)-s^*)(i(t)-i^*) -(\pi + \gamma- \beta s^*)i(t) + \beta i^* s(t) - \beta s^*i^*\, , \label{eq:SIRS_FPexpanded_i}
\end{align}
\end{subequations}
where $s^*$ and $i^*$ are the elements of the fixed point vector $\epsilon = (s^*,i^*)$ which represents the equilibrium that is reached. This equilibrium depends uniquely on the value of $R_V$. Note that the refurbished nonlinear term vanishes at the fixed point and represents the product of the fluctuations in susceptible and in infected fractions relative to the equilibrium values. This approach captures the correct asymptotic behavior for long times provided the linear relaxation times for both $s(t)$ and $i(t)$ exist. This is the case for all  $R_V \neq 1$. For $R_V = 1$ this strategy fails because $i(t)$ is then a marginal variable in the linear system since $i'(t) =0$ at linear level, which precludes an approach to the fixed point. However, a different choice of linear operator will fix this problem. We will discuss the special (critical) case $R_V = 1$ separately. The novelty of the proposed extension of the BLUES function method to coupled nonlinear systems lies in the above judicious tailoring of the linear system, which includes both equilibria by construction.

With the calibration chosen as in \eqref{eq:SIR_FPexpanded} we proceed to identify the linear operator,
\begin{equation}
    \label{eq:matrix_ODE_linear}
    \lop\vect{X} = \vect{X}_t - A \vect{X} = \vect{\chi} + \vect{C} \delta  = \vect{\psi}
\end{equation}
where the subscript $t$ on $\vect{X}$ denotes the time derivative, $A$ is the matrix with elements, 
\begin{equation}
    \label{eq:matrix_coefficients}
    A = \begin{pmatrix}
        -(\pi+\xi-\beta i^*) & -(\xi+\beta s^*) \\
        \beta i^* & -(\pi+\gamma -\beta s^*)
    \end{pmatrix}
\end{equation}
and $\vect{\chi}$ is the vector with elements
\begin{equation}
    \label{eq:vector_source_explicit}
    \vect{\chi} = \begin{pmatrix}
        \pi(1-p) +\xi + \beta s^*i^*\\
        - \beta s^*i^* \\ 
    \end{pmatrix}\,.
\end{equation}

The (nonlinear) residual operator $\rop$ applied to the solution vector $\vect{X}$ then takes the form
\begin{equation}
    \label{eq:neg_residual}
    \rop\vect{X} = \begin{pmatrix}
        -\beta (s(t)-s^*)(i(t)-i^*)\\
        \beta (s(t)-s^*)(i(t)-i^*)
    \end{pmatrix}\, .
\end{equation}
Following the procedure outlined in Section \ref{sec:BLUES_method_system}, we construct an iteration sequence \eqref{eq:nth_order} for the solution vector $\vect{X}(t)$, i.e., 
\begin{equation}
    \label{eq:blues_iteration_sequence}
    \vect{X}^{(n)}(t) = (B\ast\vect{\phi}^{(n)})(t) = \vect{X}^{(0)}(t) + \left(B\ast\rop\vect{X}^{(n-1)}\right)(t)\, ,
\end{equation}
where $B(t)$ is taken to be the matrix Green function $G(t)$ for the linear problem defined through \eqref{eq:matrix_ODE_linear}. This $G(t)$ can be found as the inverse of the fundamental matrix of the matrix of coefficients $A$ or equivalently as the matrix exponential of $tA$ multiplied by a step function, i.e., 
\begin{equation}
    \label{eq:green}
    G(t) = \mathrm{e}^{tA}\, \Theta(t)\,.
\end{equation}
 For the disease-free equilibrium ($R_V<1$) we obtain,
\begin{equation}
    \label{eq:matrixexp_free}
    G(t) = \begin{pmatrix}
        \mathrm{e}^{-(\pi+\xi)t} & \frac{\xi + (\pi+\gamma)R_V}{(\xi-\gamma) + (\pi+\gamma)R_V}\left(\me^{-(\pi+\xi)t} - \me^{-(\pi+\gamma)(1-R_V)t}\right) \\
        0 & \mathrm{e}^{-(\pi+\gamma)(1-R_V)t}
    \end{pmatrix} \Theta(t).
\end{equation}
In this (simple) case the Green function matrix is triangular, so that its eigenvalues are conspicuous on the main diagonal. These eigenvalues contain the essential ``damping" by virtue of the decaying exponentials, with finite ``linear relaxation times" $\tau_s = 1/(\pi+\xi)$ and $\tau_i = 1/((\pi+\gamma)(1-R_V))$. Note that $\tau_i$ diverges for $R_V \uparrow 1$. In this limit the relaxation to the disease-free equilibrium becomes ``nonlinear". The damping (for $R_V <1$) ensures that the long-time asymptotics of the approximants, calculated through convolution, are well behaved. 

The general Green function matrix, appropriate for both disease-free and endemic equilibria, is more involved and reads,
\begin{equation}
    \label{eq:matrixexp_general}
    G(t) = \frac{\me^{-\frac{Lt}{2}}}{2M}\begin{pmatrix}
        Z_+\,\me^{\frac{Mt}{2}} +Z_-\,\me^{-\frac{Mt}{2}} & 2\left(\me^{-\frac{Mt}{2}} -\me^{\frac{Mt}{2}}\right)(\beta s^*+\xi) \\
        2\left(\me^{\frac{Mt}{2}} -\me^{-\frac{Mt}{2}}\right)\beta i^* &  Z_+\,\me^{\frac{-Mt}{2}} + Z_-\,\me^{\frac{Mt}{2}}
    \end{pmatrix} \Theta(t)\,,
\end{equation}
with 
\begin{equation}
    \begin{split}
        Z_{\pm} &= M \pm K\\
        K &= \gamma-\xi-\beta(s^* + i^*)\\
        L &= \gamma + \xi +2\pi -\beta(s^* -i^*)\\
        M^2 &= \gamma^2 + \left[\beta(s^*-i^*) +\xi\right]^2 -2\gamma(\gamma-K)\,.
    \end{split}
\end{equation}

The zeroth approximant \eqref{eq:zeroth_order} is the convolution  of the matrix Green function with the source vector $\vect{\psi}(t)$, i.e.,
\begin{equation}
    \label{eq:convolution}
    \begin{split}
    \vect{X}^{(0)}(t) &= (G\ast\vect{\psi})(t) = \int_\mathbb{R} G(t-t') \left[\vect{\chi} + \vect{C}\delta(t')\right]\mathrm{d}t' \\
    \end{split}
\end{equation}
and results in the following expressions for the population fraction of susceptible and infected individuals, respectively,
\begin{align}
        s^{(0)}(t) &= \frac{i_0 (\beta s^* +\xi)}{M}\me^{-L t/2}\left(\me^{-Mt/2} -\me^{Mt/2}\right)
        + \frac{s_0}{2M}\me^{-Lt/2} \left(Z_-\,\me^{-Mt/2} +Z_+\,\me^{Mt/2}\right) \nonumber \\
        &+ \frac{2\beta s^* i^* (\beta s^* +\xi)}{M}\left(\frac{(\me^{-Lt/2} - \me^{-Mt/2})}{L-M} -\frac{(\me^{-Lt/2} - \me^{Mt/2})}{L+M}\right) \nonumber \\
        &+ \frac{(\beta s^* i^* + \pi(1-p)+\xi)\me^{-Lt/2}}{M}\left(\frac{Z_-\,(\me^{Lt/2} - \me^{-Mt/2})}{L+M}\right) \nonumber \\
        &+\frac{(\beta s^* i^* + \pi(1-p)+\xi)\me^{-Lt/2}}{M}\left(\frac{Z_+\,(\me^{Lt/2} - \me^{Mt/2})}{L-M}\right) \label{eq:zeroth_order_full_s}
\end{align}
\begin{align}
        i^{(0)}(t) &= \frac{s_0 \beta i^*}{M} \me^{-L t/2} \left(\me^{Mt/2} -\me^{-Mt/2}\right)
        + \frac{i_0}{2M}\me^{-Lt/2} \left(Z_-\,\me^{Mt/2} +Z_+\,\me^{-Mt/2}\right) \nonumber\\
        &- \frac{2\beta i^* (\beta s^*i^* +(1-p)\pi +\xi)\me^{Mt/2}}{M}\left(\frac{(\me^{-Lt/2} - \me^{-Mt/2})}{L-M}\right) \nonumber\\
        &+\frac{2\beta i^* (\beta s^*i^* +(1-p)\pi +\xi)\me^{Mt/2}}{M}\left(\frac{(\me^{-Lt/2} - \me^{-Mt/2})\me^{-Mt}}{L+M}\right)\nonumber\\
        &-\frac{\beta s^*i^*}{M}\me^{-Lt/2}\left(\frac{Z_-\,(\me^{Lt/2} - \me^{Mt/2})}{L-M} +\frac{Z_+\,(\me^{Lt/2} - \me^{-Mt/2})}{L+M}\right)\,. \label{eq:zeroth_order_full_i}
\end{align}

Upon inspection of this and higher approximants (not reported analytically here) we infer that all BLUES approximants, regardless of the number of iterations ($n \geq 0$), are qualitatively correct asymptotically, for all $R_V \neq 1$, in that they converge exponentially rapidly towards the exact fixed point values for long times, in contrast with the other methods which yield divergences. 

We proceed to compare graphically the solution of the SIRS model calculated with the BLUES method with a precise numerical solution and with approximate solutions obtained by the ADM, the VIM, or homotopy perturbation method (HPM). In Table \ref{table:parameters} the parameters are shown for three  different cases, together with the values for the vaccination reproduction number $R_V$ and critical vaccination threshold $p_c$. Depending on the value of $R_V$, we indicate in the last column of Table \ref{table:parameters} the equilibrium attained by the system.

\setlength{\tabcolsep}{5pt}
\renewcommand{\arraystretch}{1.2}
\begin{table}[!htp]
\caption{Parameters and corresponding equilibria for the 3 studied cases in the SIRS model. The vaccination reproduction number $R_V$ and critical vaccination threshold $p_c$ are also shown. Note that the latter exceeds unity in Case 2, which is physically equivalent to setting it equal to unity. For all cases $s_0 = 0.8$, $i_0 = 0.2$, $\beta = 0.8$, $\gamma = 0.03$ and $\pi = 0.4$.}
\label{table:parameters}
\begin{center}
 \begin{tabular}{|c ||  c c| c |c | c | c | c |} 
 \hline
  &  $\xi$ & $p$ & $p_c$ & $R_V$ & Equilibrium\\
 \hline\hline
  Case 1 & 0.1 & 0.9 & 0.5781 & 0.5209 & $\varepsilon_0 = (0.28,0)$\\
 \hline
 Case 2  & 0.5 & 0.9 & 1.0406 $\implies$ 1& 1.1163 & $\varepsilon_e = (0.5375, 0.0605)$\\
 \hline
Case 3  & 0.1 & 0.5781 & 0.5781 & 1 & $\varepsilon_0 =(0.5375,0) $\\
\hline\hline
\end{tabular}
\end{center}
\end{table}

\subsection*{Case 1: small loss of immunity and high vaccination probability.}\label{subsec:Case1}
\begin{figure}[!ht]
    \centering
    \includegraphics[width=0.8\linewidth]{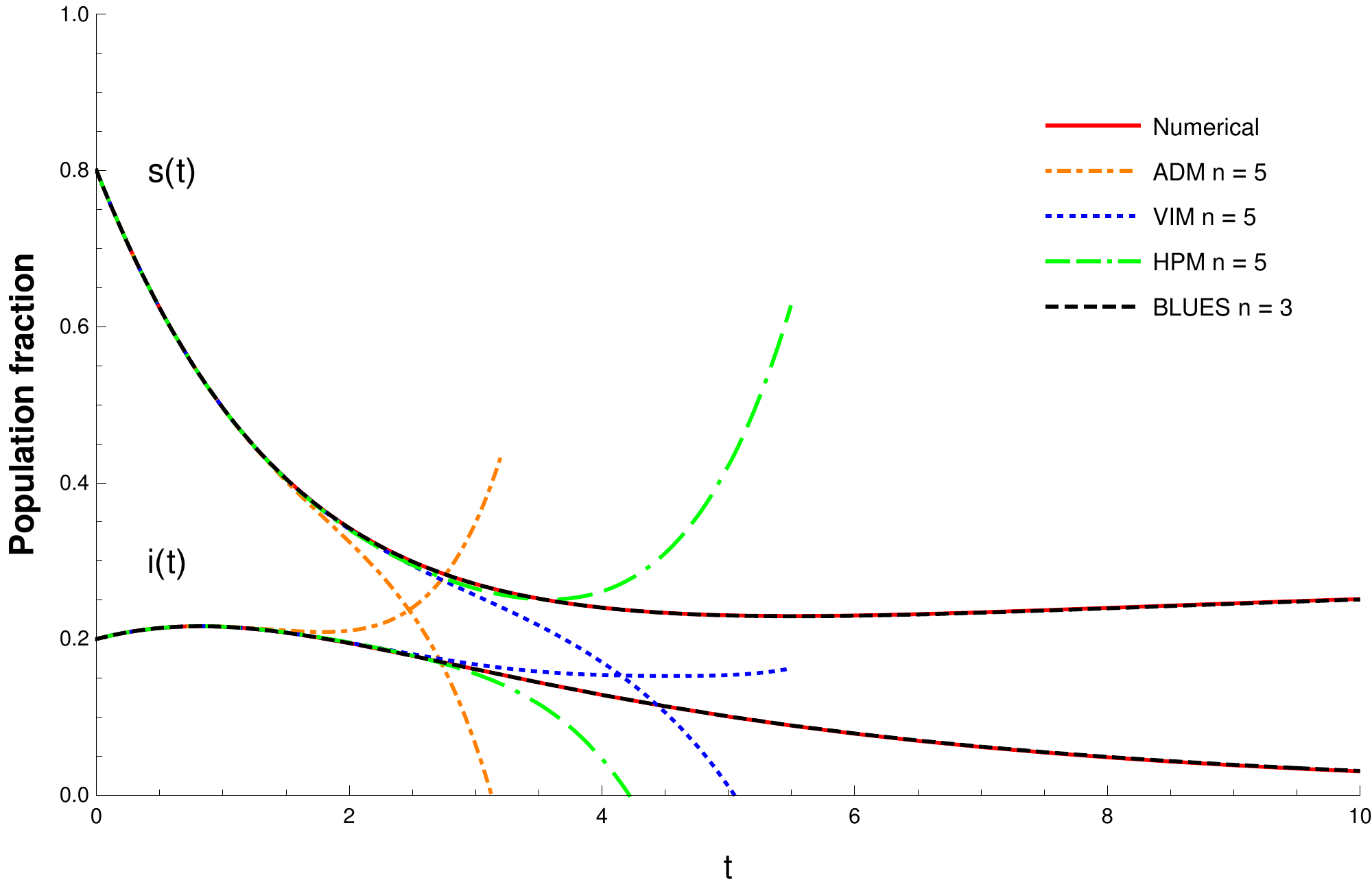}
    \caption{Comparison between the numerical solution (red line), the fifth-order ADM (orange dot-dashed line), the fifth-order VIM (blue dotted line), the fifth-order HPM (green dot-dash-dashed line) and the third BLUES approximant  (black, dashed line) for Case 1 of the SIRS model: $\xi= 0.1$ and $p=0.9$. In this case the disease-free equilibrium is reached for $t \rightarrow \infty$. Note that the numerical solution and the BLUES approximant are indistinguishable at this resolution. }
    \label{fig:SIRS_Case1}
\end{figure}

As a preliminary remark, we mention that for $\xi =0$ (no loss of immunity) the SIRS reduces to a SIR model with vaccination, which was treated earlier in \cite{MAKINDE2007842,ghotbi2011} by means of the ADM, HPM and VIM.
Here we consider the SIRS model in which the protection offered by vaccination or post-disease immunity is lost with a small probability $\xi = 0.1$ after some time. When the vaccination probability $p = 0.9$ is higher than the critical vaccination threshold $p_c = 0.5781$, the disease will eventually die out and the system will reach the stable disease-free equilibrium for which $i\rightarrow0$ and $s\rightarrow0.28$. This is shown in Fig. \ref{fig:SIRS_Case1}.  As we already discussed the BLUES method is accurate and captures the fixed-point values \eqref{eq:fixed_points_0} of $(s^*_0, i^*_0)$ in the equilibrium exactly. We remark that the approximants generated by the ADM, VIM  and HPM diverge uncontrollably for longer times while  the BLUES approximants converge globally for all $t\geq 0$ and in every iteration.

\begin{figure}[!ht]
    \centering
    \includegraphics[width=0.8\linewidth]{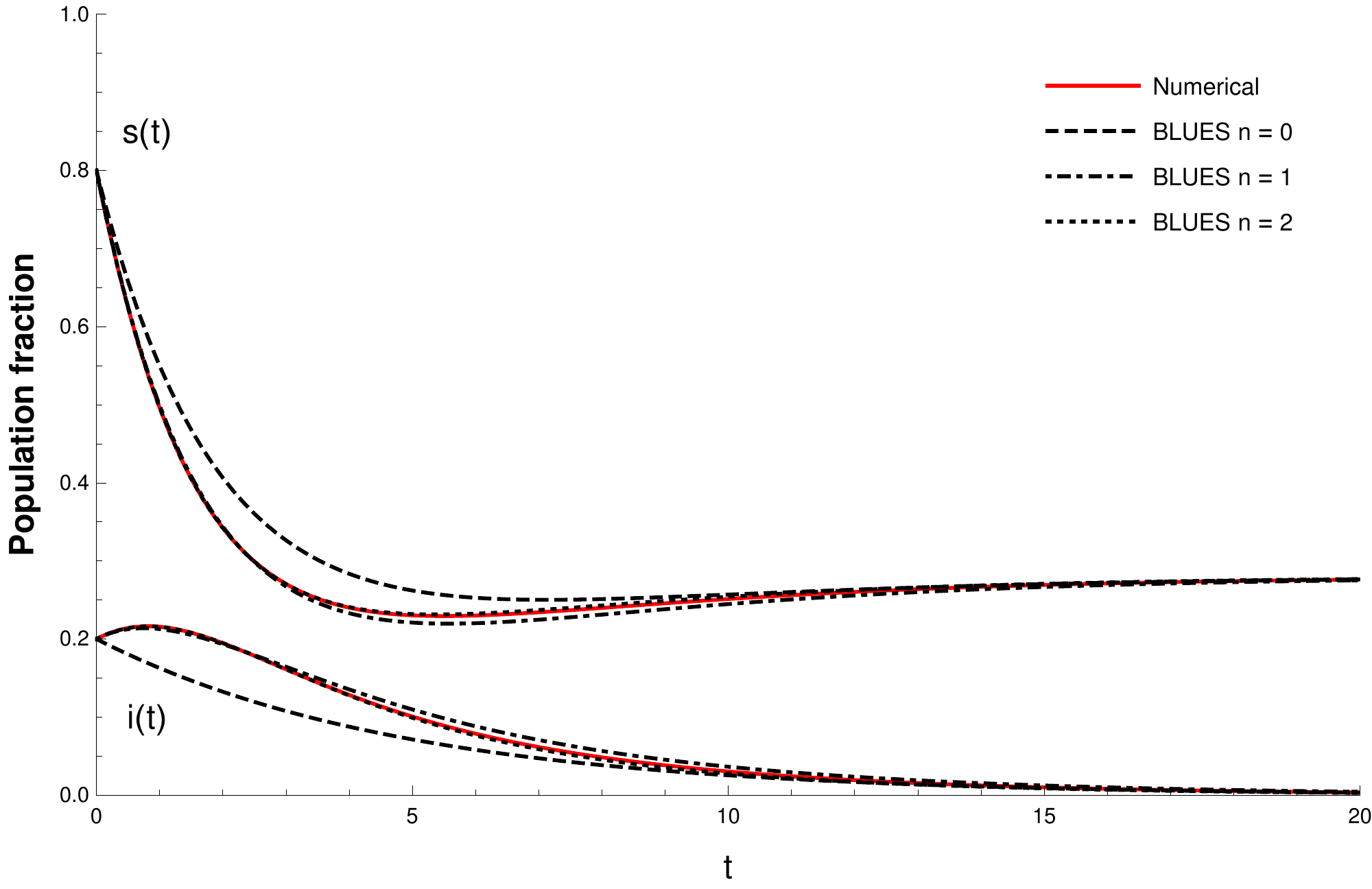}
    \caption{Comparison and convergence of BLUES approximants after zero (dashed line), one (dot-dashed line) and two (dotted line) iterations. The numerical solution is also shown (red line). This figure is for Case 1 of the SIRS model: $\xi= 0.1$ and $p=0.9$. The convergence is global and all approximants (for all $n$) attain the exact fixed point values for $t \rightarrow \infty$. The asymptotic behavior for long times is an exponential decay. Note that the numerical solution and the second BLUES approximant are
indistinguishable at this resolution. }
    \label{fig:SIRS_Case1_BLUEScompar}
\end{figure}

We also compare the BLUES approximants for different numbers of iteration and notice that they converge rapidly towards the numerical solution. This is shown in Fig. \ref{fig:SIRS_Case1_BLUEScompar}. 

\subsection*{Case 2: high loss of immunity and high vaccination probability.}\label{subsec:Case2}
\begin{figure}[!htp]
    \centering
    \includegraphics[width=0.8\linewidth]{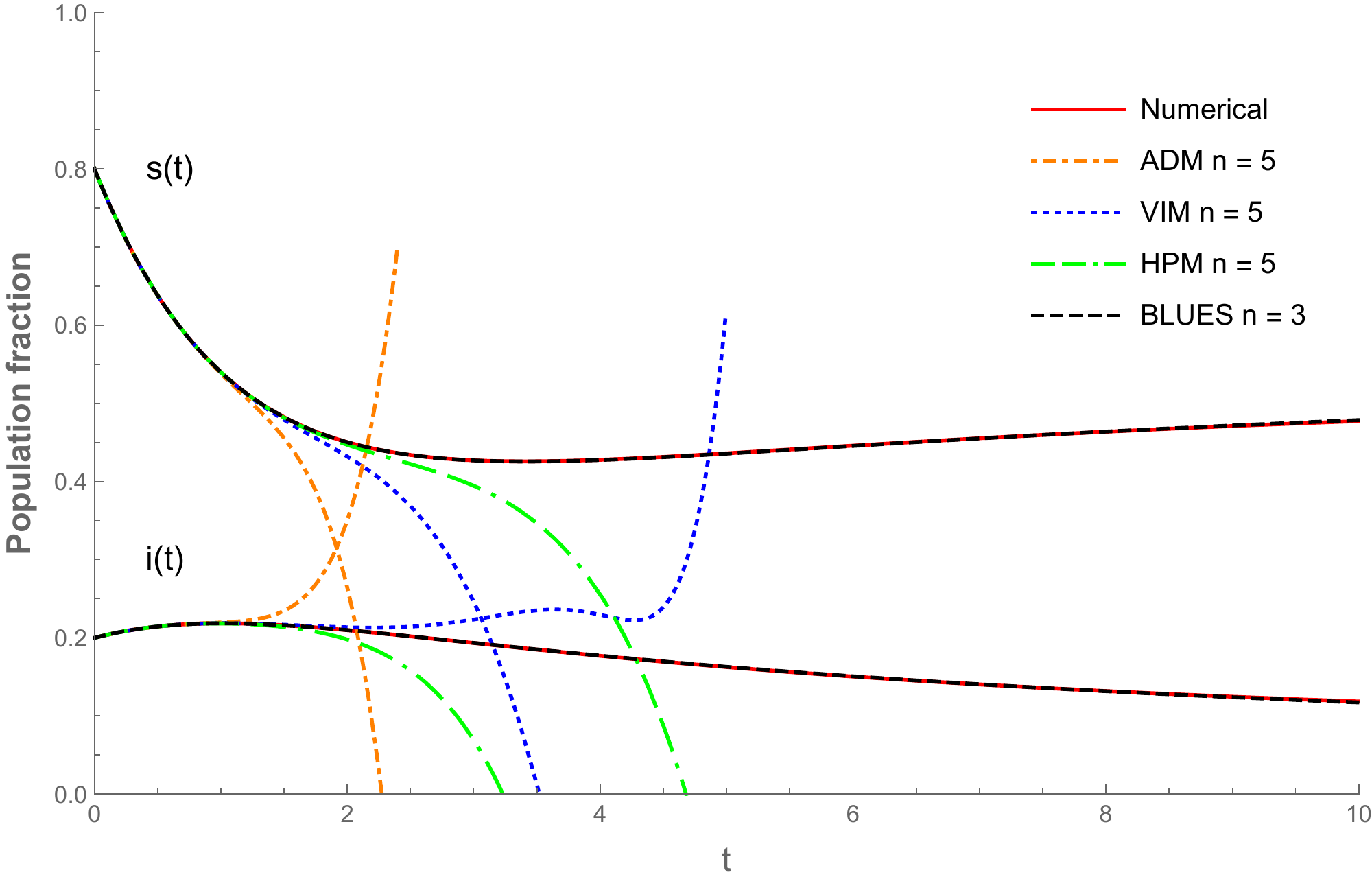}
    \caption{Comparison between the numerical solution (red line),  the fifth-order ADM, VIM, and HPM approximants (respectively, blue dotted, orange dot-dashed and green dot-dash-dashed lines) and the third BLUES approximant (black, dashed line) for Case 2 of the SIRS model: $\xi =0.5$ and $p=0.9$. In this case the endemic equilibrium is reached for $t \rightarrow \infty$. Note that the numerical solution and the BLUES approximant are indistinguishable at this resolution. }
    \label{fig:SIRS_Case2}
\end{figure}

\begin{figure}[!htp]
    \centering
    \includegraphics[width=0.8\linewidth]{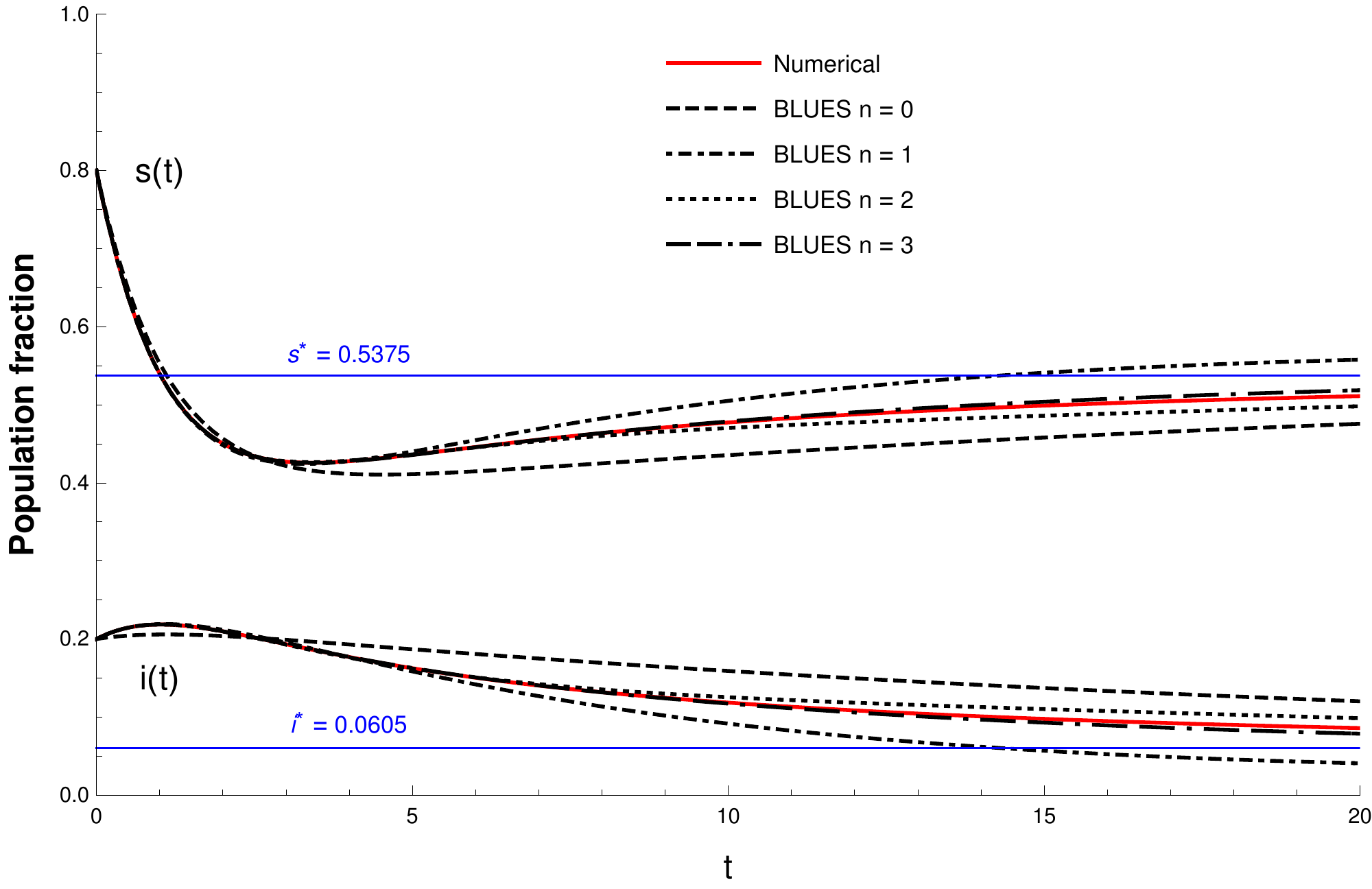}
    \caption{Comparison and convergence of BLUES approximants after zero (dashed line), one (dot-dashed line), two (dotted line)  and three (dot-dash-dashed line) iterations. The numerical solution is also shown (red line).  This figure is for Case 2 of the SIRS model: $\xi =0.5$ and $p=0.9$. The convergence is global and all approximants (for all $n$) attain the exact fixed point values for $t \rightarrow \infty$ indicated by the horizontal blue lines. The asymptotic behavior for long times is an exponential decay. }
    \label{fig:SIRS_Case2_BLUEScompar}
\end{figure}

As a  second example, we consider the case in which immunity is more easily lost $(\xi=0.5)$ and the population is putting in an effort to vaccinate a larger number of people $(p=0.9)$. We can deduce from the critical vaccination probability $p_c = 1.0406$ in Table \ref{table:parameters} that even when all civilians are vaccinated, immunity is lost so quickly that the population always reaches the endemic equilibrium and the disease cannot be eradicated. The result of a comparison between the ADM, VIM,  HPM and BLUES methods is shown in Fig. \ref{fig:SIRS_Case2}. We also compare the BLUES approximants for different numbers of iteration and observe that they converge rapidly towards the numerical solution. This is shown in Fig. \ref{fig:SIRS_Case2_BLUEScompar}.

\subsection*{Case 3: nonlinear relaxation at the dynamical critical point.}\label{subsec:Case3}
In this third case, we study the dynamical criticality at $R_V = 1$. The population still reaches the disease-free equilibrium asymptotically, but much more slowly since in the limit $R_V \uparrow 1$ the linear relaxation time diverges. The ``linear" exponential relaxation is replaced by a ``nonlinear" algebraic one, with leading behavior for long times proportional to $1/t$. This can be inferred exactly from an analysis of the asymptotic behavior of the system of DEs \eqref{eq:SIR_FPexpanded}, which at $R_V=1$ reduces to,
\begin{subequations}
\label{eq:SIR_FPexpanded_critical}
\begin{align}
s'(t) &=  \pi(1-p) -\beta (s(t)-s^*)i(t) - (\pi+\xi) s(t) -(\xi+\beta s^*) i(t) +\xi  \label{eq:SIRS_FPexpanded_s_c}\\
i'(t) & = \beta (s(t)-s^*)i(t) \, , \label{eq:SIRS_FPexpanded_i_c}
\end{align}
\end{subequations}
with, in this special case, $s^* = s^*_0=s^*_e$. Inspection of these DEs allows us to establish that the leading asymptotic behavior is a $1/t$ power-law decay towards the fixed point,
\begin{subequations}
\label{eq:algebraicasymptotics}
\begin{align}
s(t) &=  s^* - \frac{1}{\beta \, t} + {\cal O}(t^{-2})  \label{eq:asymp_s}\\
i(t) & =  \frac{\pi+\xi}{\pi+\xi +\gamma} \,\frac{1}{\beta \, t} + {\cal O}(t^{-2}) \label{asymp_i},
\end{align}
\end{subequations}

A successful application of the BLUES function method is possible if we acknowledge that we must reconsider the decomposition of the problem into a nonlinear and a linear part. This is necessary in view of the divergence of the ``linear relaxation time" $\tau_i = 1/((\pi+\gamma)(1-R_V))$ and the concomitant vanishing of one of the eigenvalues of the linear matrix $A$ given by the triangular form \eqref{eq:matrix_coefficients}, which is appropriate for the disease-free equilibrium. If we stick to this choice there is not enough ``damping" in the convolution products that govern the iteration procedure \eqref{eq:nth_order} and the approximants diverge, similarly to what routinely happens in the ADM, the VIM and HPM. However, we can recalibrate the linear operator so that both eigenvalues are non-zero and nevertheless the correct (disease-free) fixed point is reached for $t \rightarrow \infty$. This is achieved simply by retaining the original nonlinear term in \eqref{eq:SIR_normalized_invariant} without refurbishing it. The matrix of coefficients $A$ of the linear operator and the residual operator hence become, respectively,

\begin{equation}
    \label{eq:A_matrix_R1}
    A = \begin{pmatrix}
         -\pi -\xi  & -\xi  \\
         0 & -\pi-\gamma  \\
    \end{pmatrix}
\end{equation}
and 
\begin{equation}
    \label{eq:neg_residual_R1}
    \rop\vect{X} = \begin{pmatrix}
        -\beta s(t) i(t)\\
        \beta s(t) i(t)
    \end{pmatrix}\,.
\end{equation}

\begin{figure}[!ht]
    \centering
    \includegraphics[width=0.8\linewidth]{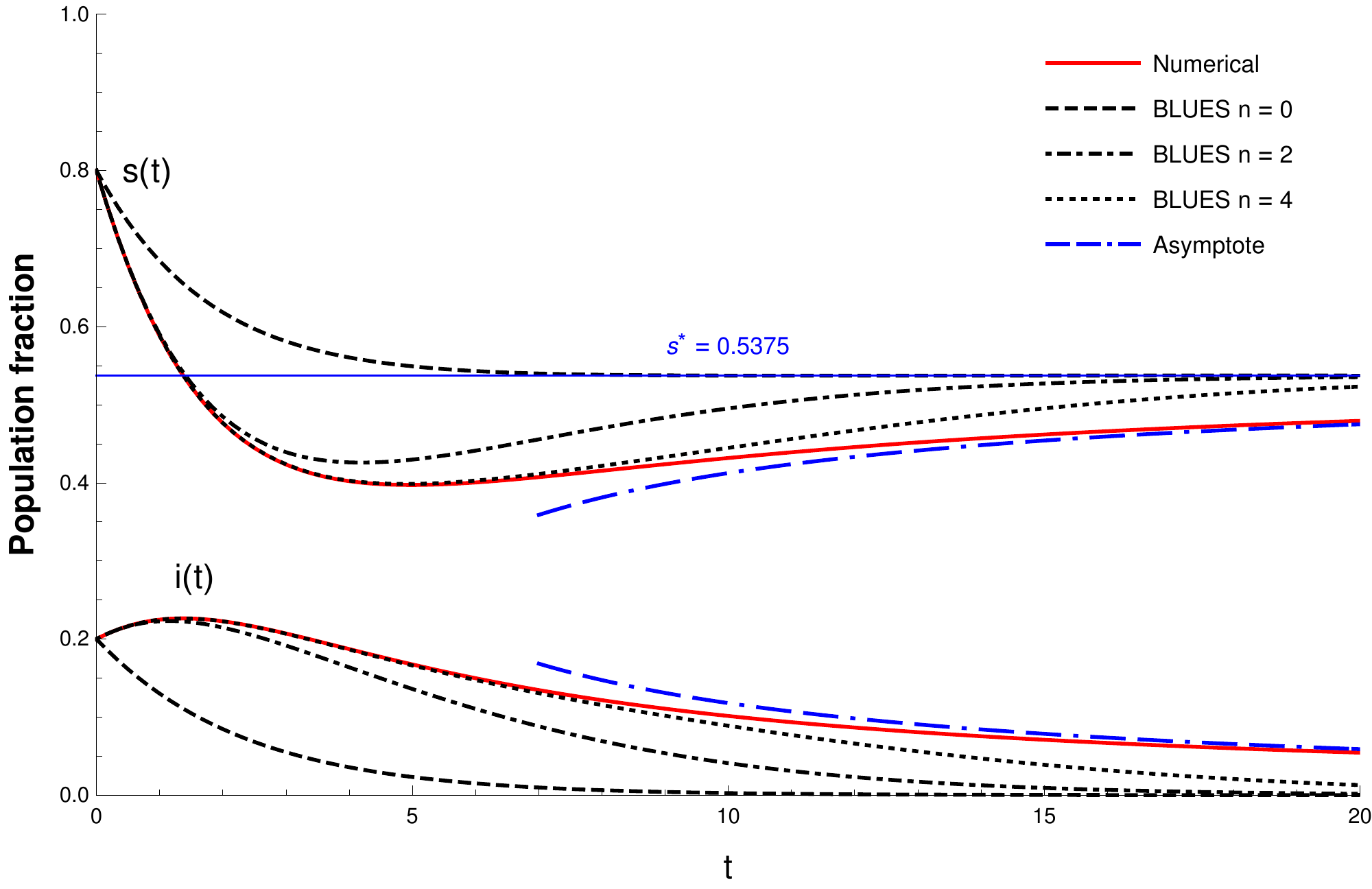}
    \caption{Comparison and convergence of the BLUES approximants after zero (dashed line), two (dot-dashed line) and four (dotted line) iterations. The numerical solution is also shown (red line) together with its exact asymptotic behavior (a power-law decay). This figure is for Case 3 of the SIRS model: a critical point with $ \xi = 0.1$ and $p = p_c = 0.5781$, implying $R_V=1$. The convergence is global and all approximants (for all $n$) attain the exact fixed point values for $t \rightarrow \infty$. However, the asymptotic behavior of the exact solution for long times is an algebraic decay, whereas the BLUES approximants feature an exponential decay.}
    \label{fig:SIRS_Case3_BLUEScompar}
\end{figure}

This linear system ensures a correct approach to the fixed point for $R_V=1$ and is a good starting point for the BLUES iteration in this dynamical critical point. We recover global convergence to the numerically exact solution, but at a slower pace than in the noncritical case because the approximants decay exponentially fast for long times, whereas the exact solution features an algebraic decay. The characteristic time of the leading exponentials is constant for all approximants but the amplitude, sign and polynomial prefactors vary as the iteration number $n$ increases. This is how the iteration sequence attempts to approximate an algebraic decay in the limit $n \rightarrow \infty$ (see Fig.\ref{fig:SIRS_Case3_BLUEScompar}).
 
\begin{figure}[!ht]
    \centering
    \includegraphics[width=0.8\linewidth]{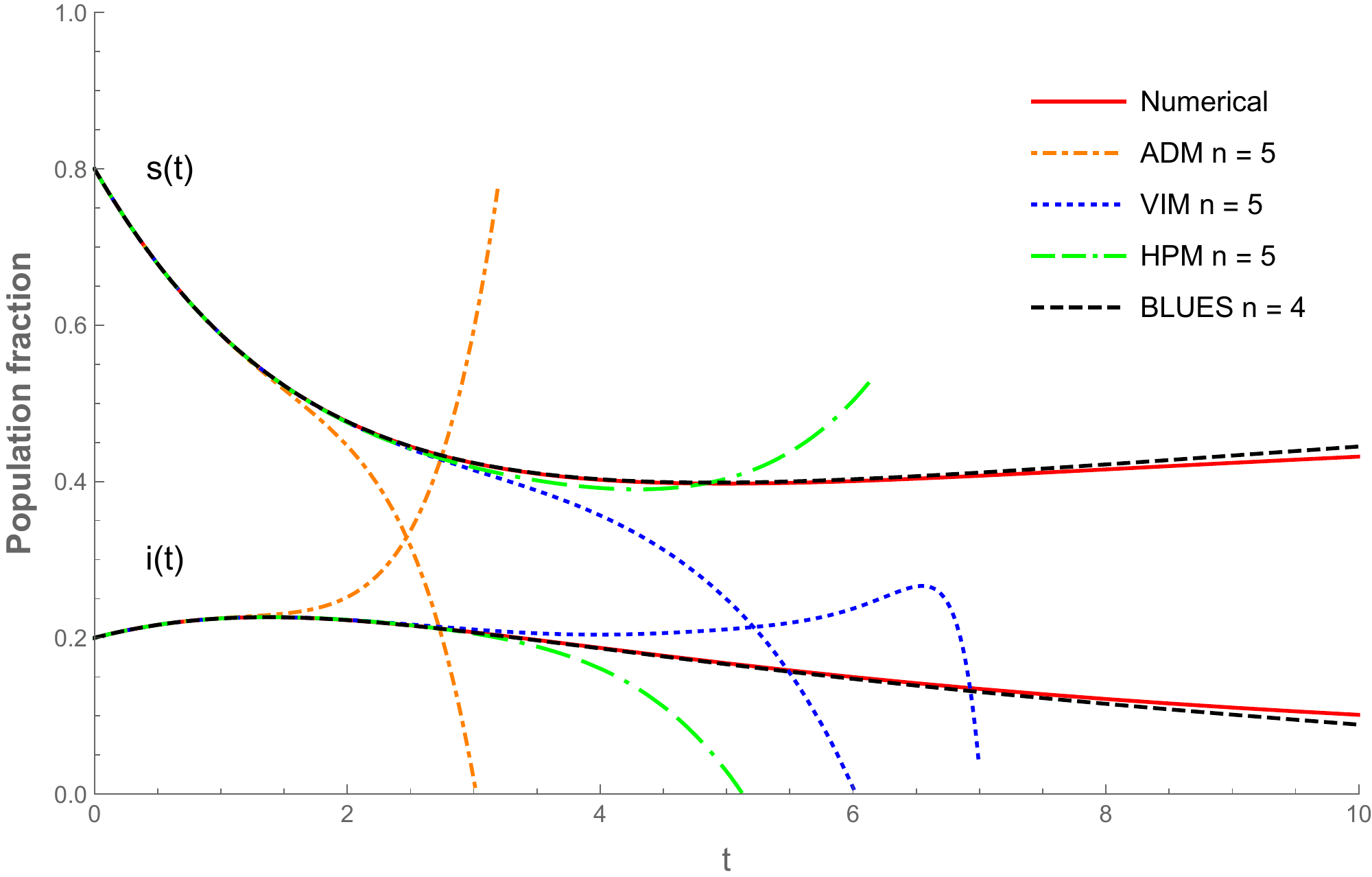}
    \caption{Comparison between the numerical solution (red line),  the fifth-order ADM, VIM, and HPM approximants (respectively, blue dotted, orange dot-dashed and green dot-dash-dashed lines) and the fourth BLUES approximant (black, dashed line). This figure is for Case 3 of the SIRS model: a critical point with $\xi = 0.1$ and $p = p_c = 0.5781$, implying $R_V=1$. Note that the BLUES approximant is initially very close to, but later deviates somewhat from the numerical solution. This is due to the difference in the type of asymptotic decay, which is conspicuous in Fig.\ref{fig:SIRS_Case3_BLUEScompar}.}
    \label{fig:SIRS_Case3}
\end{figure}

\subsection*{Time of the infection peak}\label{Subsec:peaktime}
\begin{figure}[!htp]
    \centering
    \includegraphics[width=0.8\textwidth]{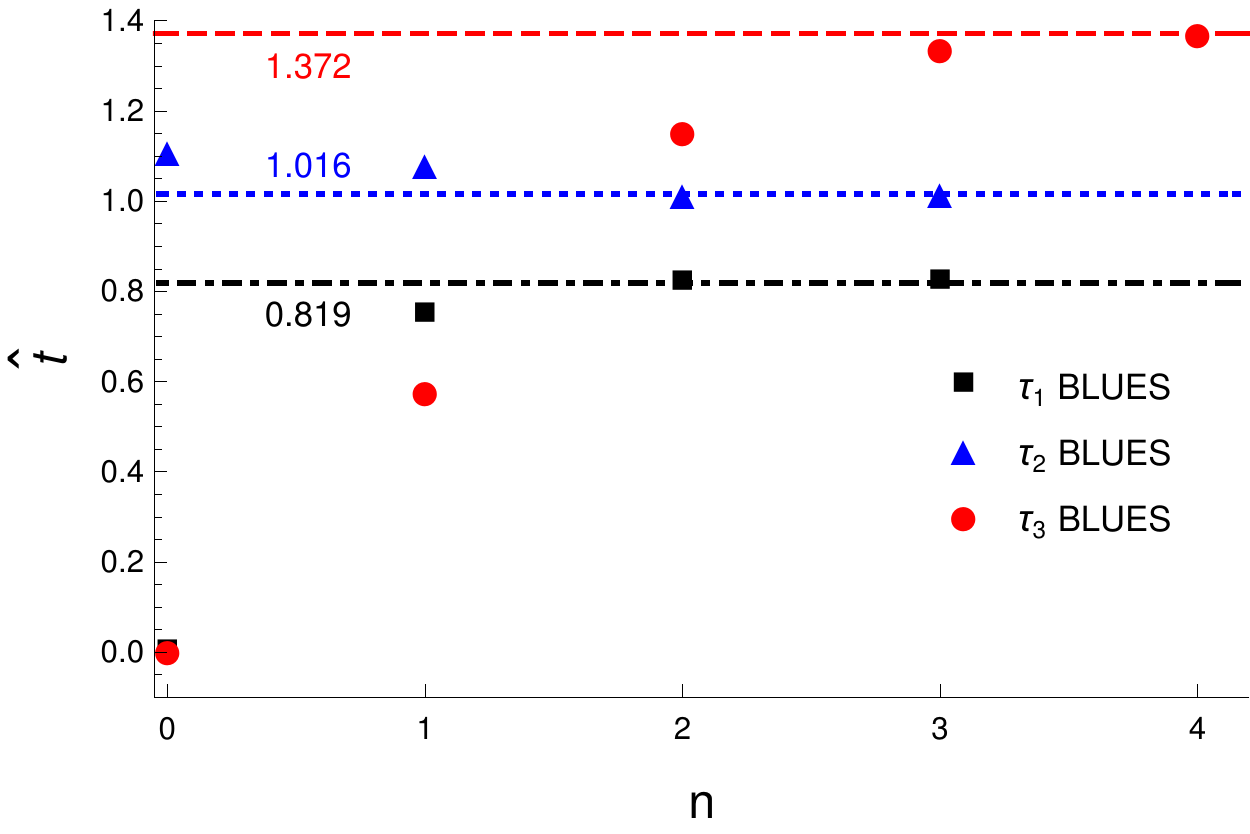}
    \caption{Comparison between the numerically precise (dimensionless) time of the infection peak $\hat t$  in the SIRS model and the values calculated using the $n$th BLUES approximants ($n=0,1,2,3,4$)  for Case 1 (red dashed line and red circles), Case 2 (black dot-dashed line and black squares) and Case 3 (blue dotted line and blue triangles).}
    \label{fig:maximuminfectionplot}
\end{figure}

We now calculate the (dimensionless) time $t=\hat t$ at which the peak of the infection occurs from the BLUES approximants and compare with the numerically precise values. At the infection peak, $i'(\hat t) = 0$, and hence, using equation \eqref{eq:SIRS_invariant_i}, we deduce that $s(\hat t) = (\gamma+\pi)/\beta$. So, instead of trying to solve $i'(\hat t) = 0$ directly, we can find $\hat t$ from the susceptible population fraction. The results are shown in Fig. \ref{fig:maximuminfectionplot}. The BLUES function method accurately captures the infection peak time, both for the disease-free and the endemic equilibria.

\section{Conclusions}
\label{sec:conclusions}
In this paper we have presented a twofold advance. The first consists of qualitative progress in accuracy and convergence of analytic approximations to solutions of the SIRS model for epidemic spreading. The second is concerned with a technical and methodological refinement, being the development of a {\em matrix} BLUES function method for {\em coupled} nonlinear ordinary differential equations. 

We have shown that the method can be applied to obtain analytic approximants for the SIRS model with vital dynamics, constant vaccination strategy and loss of immunity. We have made a detailed comparison of the iteration procedure with the Adomian decomposition method, the variational iteration method and homotopy perturbation method. It is found that all methods succeed in approximating the (numerically) exact solutions locally and, for the BLUES function method, also globally. The BLUES method generates approximants with a higher accuracy, for the same number of iterations performed, and, moreover, is able to capture the asymptotic behavior of the solutions for long times in both the disease-free and endemic equilibria. In unpublished work \cite{Berx_thesis} we have also studied the SEIRS model, which is an extension of the model treated in this paper.

A prominent strength of the BLUES function method is that it permits the user to tailor judiciously the linear part of the operator so that its Green function contains just enough damping to temper, in each iteration, its convolution with the emergent source in order to capture the correct asymptotics of the solution of the nonlinear problem. In other words, for an optimal choice of the linear part and its associated Green function, the BLUES iterations converge globally to the numerically exact solution of the problem. This choice can be viewed as a guiding principle in choosing the associated linear operator. For a suboptimal choice of the linear part local convergence can still be achieved, but in that case the advantages over other methods are less pronounced. An illustration of an optimal and suboptimal choice of linear operator has been presented in the case of the dynamical critical point of the SIRS model. In general, when using the BLUES function method to generate analytical approximants to systems of coupled DEs, one should tailor the linear system in such a way that it includes all of the existing steady states and at the same time respects the initial condition by means of a source term, in order to achieve globally convergent results.

With the ongoing COVID-19 pandemic, it has become clear that mathematical modeling of epidemic processes is crucial to understanding (and possibly also predicting) the evolution of these viral outbreaks. None of the epidemiological models have exact closed-form solutions, with the exception of the SIS model \cite{Hethcote1989} in which no immunity is possible and individuals return to the susceptible contingent once they recover from the infection. Hence, approximate solutions are needed to assess the impact of the different model parameters without resorting to brute force methods such as numerical simulation. This paper has aimed at showing that the BLUES method is a good candidate for obtaining such approximate analytic solutions. This application of the method to coupled ODEs paves the way for applications to more involved systems such as coupled PDEs \cite{Berx2021_PDE,Horowitz}, for instance.  
\appendix 

\section{Stability analysis}
\label{sec:stability}
The system \eqref{eq:SIR_normalized_invariant} has two fixed points that can be found by a fixed-point analysis which reveals a disease-free equilibrium $\varepsilon_0$ in which the disease has died out, and an endemic equilibrium $\varepsilon_e$ in which the infected population density reaches a nonzero asymptotic value, i.e., 
\begin{subequations}
\label{eq:SIRS_equilibria}
\begin{align}
    \varepsilon_0 &= (s^*_0,i^*_0) =\left(1-\frac{\pi p}{\pi +\xi},0\right) \label{eq:fixed_points_0}\\
    \varepsilon_e &= (s^*_e,i^*_e) = \left(\frac{\pi+\gamma}{\beta},\frac{\beta((1-p)\pi +\xi) -(\gamma+\pi)(\xi+\pi)}{\beta(\gamma+\pi+\xi)}\right). \label{eq:fixed_points_e_full}
\end{align}
\end{subequations}

Note that the disease-free equilibrium $\varepsilon_0$ is independent of the average contact rate $\beta$. The endemic equilibrium \eqref{eq:fixed_points_e_full} can now be simplified to
\begin{gather}    
    \varepsilon_e = (s^*_e,i^*_e) = \left(\left(1-\frac{\pi p}{\pi +\xi}\right)\frac{1}{R^{({\rm SIRS})}_V},\frac{(1-p)\pi+\xi}{\gamma+\pi+\xi}\left(1-\frac{1}{R^{({\rm SIRS})}_V}\right)\right). \label{eq:fixed_points_e}
\end{gather}
These two fixed points are globally stable. This means that for an arbitrary initial condition in the $si$-plane, the trajectory will converge onto one of these two fixed points. Which fixed point is reached depends on the system parameters $\pi,\beta,\gamma,\xi$ and $p$.

The global asymptotic stability for the endemic equilibrium can be proven as follows. First note that the positive quadrant $\mathbb{R}^2_+$ of the $si$-plane is not an invariant set of the system \eqref{eq:SIR_normalized_invariant}, i.e., when $s(t)=0$ then $s'(t)<0$ for all values $i(t)>(\pi(1-p) +\xi)/\xi$. This can be resolved by shifting $(s,i)$ to $(\Sigma,i)$, where
\begin{equation}
    \label{eq:sigma}
    \Sigma(t) = s(t) + \frac{\xi}{\beta}.
\end{equation}
Hence, the shifted system becomes
\begin{subequations}
\label{eq:SIR_normalized_shifted}
\begin{align}
\Sigma'(t) &=  \pi(1-p)-\beta \Sigma(t) i(t) -(\pi+\xi) \Sigma(t) + \frac{\xi}{\beta}(\xi+\pi)+\xi \label{eq:SIRS_shifted_sigma}\\
i'(t) & = \beta \Sigma(t)i(t)-(\gamma+\pi+\xi)i(t)\, . \label{eq:SIRS_shifted_i}
\end{align}
\end{subequations}
It is now easy to see for $\Sigma(t)=0$, now $\Sigma'(t)\geq 0$ for all values of $i(t)$. By shifting the system, the endemic equilibrium coordinates change as follows
\begin{equation}
    \label{eq:endemic_shifted}
    (\Sigma^*_e,i^*_e) = \left(\frac{\pi+\gamma+\xi}{\beta},\frac{\pi ((1-p)\beta-\gamma-\pi) +\xi(\beta-\gamma-\pi)}{\beta(\pi+\gamma+\xi)}\right).
\end{equation}
All global properties of the system remain invariant under the shifting of the coordinates, so we can now try to find a Lyapunov function $V(\Sigma,i)$ to prove global asymptotic stability of the endemic fixed point \eqref{eq:endemic_shifted} of the shifted system \eqref{eq:SIR_normalized_shifted} and hence also of the original system \eqref{eq:SIR_normalized_invariant}. We can choose the following Lyapunov function \cite{KOROBEINIKOV2002955}
\begin{equation}
    \label{eq:lyapunov}
    \begin{split}
        V(\Sigma,i) &= \Sigma^*_e\left(\frac{\Sigma}{\Sigma^*_e}-\ln{\frac{\Sigma}{\Sigma^*_e}}\right) + i^*_e\left(\frac{i}{i^*_e}-\ln{\frac{i}{i^*_e}}\right)
    \end{split}
\end{equation}
with time derivative
\begin{equation}
    \label{eq:lyapunov_derivative}
    \begin{split}
        V'(\Sigma,i) &= \pdv{V}{\Sigma}\Sigma'(t) + \pdv{V}{i}i'(t)\\
        &= \left(1-\frac{\Sigma^*_e}{\Sigma(t)}\right)\Sigma'(t)+\left(1-\frac{i^*_e}{i(t)}\right)i'(t).
    \end{split}
\end{equation}
Now, from \eqref{eq:SIRS_shifted_sigma} and \eqref{eq:SIRS_shifted_i} it is clear that for the endemic equilibrium the following holds
\begin{equation}
    \label{eq:endemic_shifted_FP_property}
    \begin{split}
        \beta\Sigma^*_e i^*_e &= \pi (1-p)-(\pi+\xi)\Sigma^*_e + \frac{\xi}{\beta}(\xi+\pi)+\xi\\
        &= (\gamma+\pi+\xi)i^*_e\, .
    \end{split}
\end{equation}
Substituting this property into \eqref{eq:lyapunov_derivative} and simplifying gives after some algebra
\begin{equation}
    \label{eq:lyapunov_negative}
    \begin{split}
        V'(\Sigma,i) &= \left(1-\frac{\Sigma^*_e}{\Sigma(t)}\right)\Sigma'(t)+\left(1-\frac{i^*_e}{i(t)}\right)i'(t)\\
        &= -\left(\pi(1-p) + \frac{\xi}{\beta}(\xi+\pi) +\xi\right)\left(\frac{\Sigma(t)}{\Sigma^*_e}\right)\left(1-\frac{\Sigma^*_e}{\Sigma(t)}\right)^2\\
        &\leq0\,,
    \end{split}
\end{equation}
for all values of $\Sigma, i\geq0$. This concludes the proof that the endemic equilibrium is globally asymptotically stable. Proving the global stability of the disease-free fixed point is now trivial. One can repeat the previous calculations with the Lyapunov function
\begin{equation}
    \label{eq:lyapunov_disease_free}
    \begin{split}
        V(\Sigma,i) &= \Sigma^*_0\left(\frac{\Sigma}{\Sigma^*_0}-\ln{\frac{\Sigma}{\Sigma^*_0}}\right) + i\, ,
    \end{split}
\end{equation}
which is the same as \eqref{eq:lyapunov} with now $\Sigma^*_0$ instead of $\Sigma^*_e$ and $i^*_e\rightarrow0$.

\bibliography{BLUES.bib}
\end{document}